\documentstyle[multicol,aps]{revtex}

\begin{document}
\draft

\title{Cooling of a small sample of Bose atoms with accidental degeneracy}

\author{Maciej Lewenstein$^{1}$, J. Ignacio Cirac$^{2}$, and Luis Santos$^{1}$}

\address{(1) Institut f\"ur Theoretische Physik, Universit\"at Hannover,
 D-30167 Hannover, Germany\\
(2) Institut f\"ur Theoretische Physik, Universit\"at Innsbruck,
 A--6020 Innsbruck, Austria}

%\address{(1) Commissariat \`a l'Energie Atomique, DSM/DRECAM/SPAM \\
%Centre d'Etudes de Saclay, 91191 Gif-sur-Yvette Cedex,
%France}
%\address{(2) Departamento de Fisica Aplicada, 
%Universidad de Castilla-La Mancha, 13071 Ciudad Real, Spain}

\date{\today}

\maketitle

\begin{abstract} 
A system of bosons in a harmonic trap is cooled via their 
interactions with a thermal reservoir.  We derive the master 
equation that governs the evolution of the system and may 
describe diverse physical situations: laser cooling, 
symphatetic cooling, etc. We investigate in detail the dynamics 
of the gas in the Lamb-Dicke limit, whereby the size of the 
trap is small in comparison to the de Broglie wavelength of the 
reservoir quanta. In this case, the dynamics is characterized 
by  two  time scales. First, an equilibrium is reached on a 
{\it fast time scale} within the degenerated subspaces of the 
system. Then, an equilibrium  between these subspaces is 
reached on a {\it slow time scale}.
\end{abstract}

\pacs{32.70.Jz,42.50.Fx,32.80.-t}

\begin{multicols}{2}

%\narrowtext

% -----------------------------------------------------------
\section{Introduction}
% -----------------------------------------------------------

\subsection{Bose--Einstein condensation}

The observation of effects related to the quantum statistical 
properties of weakly interacting gases of atoms\cite{ICAP} has 
become in the last decades a major challenge of atomic physics.  
Thus, a large part of the theoretical and experimental research 
has been focused during the $90$'s on 
cooling atoms confined in traps at relatively high densities 
\cite{BEC}. These studies have led to the remarkable 
experimental realisation of a Bose-Einstein condensation (BEC) in  
Rubidium \cite{Eric}, and sodium vapors \cite{wolfgang}. 
Evidence of BEC in a Lithium gas with attractive interactions has 
been also reported \cite{Bradley}. 
These remarkable achievements, have opened  fascinating possibilities 
and applications, such as the developement of 
a coherent source of atoms, or {\it atom laser}\cite{boser}.

The theoretical description of a system of ultracold 
bosonic or fermionic atoms is particularly convenient in the 
framework of second quantization. Quantum field theories of 
cold atoms \cite{lew94}, originally developed in a condensed--matter 
context, have been used in the  
diagnostics of a Bose-Einstein condensate \cite{diag}, and in 
{\it nonlinear atomic optics}\cite{nao}.  

Several theoretical works in the recent years were more directly 
aimed to the dynamics of the 
cooling processes, dynamics of the possible phase transitions 
and the formation of quantum collective states. Those works 
concern both {\it collisional  cooling} mechanisms, such as 
evaporation \cite{Hess,evap} or symphatetic cooling 
\cite{Wine1,Wine4,Wine5,Wineland}, or {\it  laser cooling} 
mechanisms, such as  sideband cooling \cite{sideband}, Raman 
cooling \cite{Davidson}, and dark state cooling \cite{Aspect}. 
The latter processes allow to reach temperatures below the 
photon recoil energy $E_R$ [equal to $(\hbar k)^2/(2M)$, where 
$k$ is the laser wavevector and $M$ the atomic mass]. One 
expects that the system might then enter a collective quantum 
state (such as Bose-Einstein condensate, or some analogue of 
it). In particular, under such conditions one hopes to realize 
also a coherent source of atoms. 
%Analogously, in the area of atomic 
%optics one studies the formation of atomic solitons etc. 
%\cite{Schern}.

\subsection{Quantum Master Equation}

In general, the quantum dynamics of a system of cold atoms  is 
a very complex many-body problem. Some of the above mentioned 
processes may be analyzed using quantum Boltzmann equations  
\cite{Stoof,Reichl}. Starting from $1994$ a more general method 
based on a 
quantum master equation (QME) description \cite{Gardinerbook} 
has been developed. In particular, a QME describing the dynamics 
of a small sample of  laser cooled  atoms in a harmonic 
microtrap has been proposed and analyzed \cite{CLZ}. The 
quantum statistical nature of the atoms is reflected in the 
dynamics of the cooling process. In the case when the trapping 
potentials for the atoms in the ground and excited  electronic 
states are different such QME might lead to multistability and 
{\it generalized Bose-Einstein distributions} \cite{gbed}.   
The QME has the advantage that it permits to study atom number 
fluctuations in each of the trap levels, and thus provides a 
more complete description of the cooling process. In 
particular, it may be used, in principle, to describe the 
dynamics of condensate formation. We have also derived a QME for 
symphatetic cooling \cite{sympa} and analyzed the possibility 
of achieving the condensation of a system  of light particles 
which interact with a reservoir of heavier particles. In the 
context of nonlinear atomic optics a self-consistent 
Born-Markov-Hartree-Fock master equation has been derived and 
analyzed to study the spontaneous emission effects on atomic 
solitons \cite{bmhf}.

Since that early works, the theory of QME for many--body systems has been 
very strongly developed. In particular, C. Gardiner, P. Zoller and 
co--workers developed in a series of papers \cite{QKpapers}
the QME and Quantum Kinetic Boltzmann approach to describe the 
dynamics of the evaporative cooling. We have also extended the theory 
of the collective laser cooling to much more realistic situations, 
avoiding the microtrap assumption, and working beyond the so--called 
Lamb--Dicke limit, in which the trap is of the size of the laser 
wavelength \cite{Dincol}. In such analysis, we avoided the reabsorption 
problem by working in the so--called {\em Festina Lente} regime, in 
which the spontaneous emission rate is smaller than the trap frequency.

One should stress that, apart from the area of quantum optics, 
master equations for quantum Bose or Fermi gases have been used 
in statistical physics \cite{kutner}. However, the master 
equation is usually postulated there starting from general 
statistical requirements. It describes the approach towards the 
thermal equilibrium described by the Bose-Einstein or, 
correspondingly, Fermi-Dirac distributions, it fullfills 
detailed balance conditions, and sometimes it conserves  some 
order parameters. The dynamics that it generates might have 
some universal properties, but does not have a direct physical 
interpretation in terms of interaction with specific energy  
reservoirs. 

In most of the quantum optical examples, the master equation is 
derived via the elimination of the ``bath'' degrees of freedom 
starting from a more general theory that describes a very 
specyfic physical situation. The  eliminated ``bath'' has a 
direct physical interpretation - it consists of photons, 
colliding atoms etc. Each of the jumps between the states of 
the system described by the QME usually corresponds  to a very 
well defined physical process of photon emission, absorption, 
atom--atom collision etc. \cite{fotka}. 

The QME approach, although valid in general, can only be used 
for practical calculations  when the QME can be reduced to a set 
of kinetic equations, while the density matrix to a diagonal 
form. Such reduction is not always possible, and very often 
requires first the choice of an appropriate basis in the 
Hilbert space of the  states of the system. In particular, the 
{\it bare} states of the system, that are the eigenstates of 
the system Hamiltonian in isolation from the ``bath'', are not 
necessarily the right ones. For instance, in the process of   
cooling of an ideal bosonic gas confined in a microtrap, the  
reduction in the basis of {\it bare} states is possible only if 
additional assumptions are made. These statements seem a little 
surprising, since they hold even for arbitrary weak 
system-``bath'' interactions, i.e. a situation in which the 
quantum Boltzmann equations should be valid \cite{Reichl}. 
Note, however, that the validity of the quantum Boltzmann 
equations usually requires  assumptions concerning quantum 
ergodicity, which simply do not hold in the above mentioned 
situations. The point here is that an ideal bosonic gas in a 
harmonic trap has plenty of degenerated {\it bare} energy 
levels. The quantum coherences between those levels (i.e. 
non-diagonal elements of the density matrix) can survive for 
very long times, and contribute significantly to the dynamics.

\subsection{Degeneracy in many--body systems}

Let us enumerate by $\vec m$ the eigenstates of {\it a single 
atom}   Hamiltonian    in the rotationally symmetric harmonic 
trap of frequency $\omega$,  where $\vec m$ is  a natural 
number in one dimension, a pair of natural numbers in 2D, a 
triple in 3D etc. When we consider an ensemble of $N$ atoms, 
the states of such {\it an  ideal gas} can be written in the 
Fock representation as $|n_{\vec 0},n_{\vec 1},\ldots\rangle$, 
where $n_{\vec m}$ denote the occupation numbers of the 
corresponding $\vec m$-th eigenstate. For noninteracting 
atoms there are two kinds of  degeneracies in such a system. 
First, a  degeneracy of energy levels due to rotational 
invariance; that is for the states for which the sum of 
$n_{\vec m}$'s with a fixed sum of the components of $\vec m$, 
is fixed itself. Obviously, such degeneracies are not present 
in 1D. We shall not discuss them here. Second, there exists an 
{\it accidental} degeneracy, due to the particular symmetry of 
the harmonic potential \cite{Watkin}. This degeneracy occurs 
even in the case of 1D: for instance for the states 
$|0,2,0,\ldots\rangle$ and $|1,0,1,0,\ldots\rangle$.  Here, the 
state with two atoms in the first energy level has an energy 
$2\times \hbar\omega$, which is equal to the energy of the 
state with one atom in the ground level and another atom in the 
second excited level ($1\times 0\hbar \omega + 1\times 
2\hbar\omega$). Both kinds of degeneracies are lifted up if one 
considers anisotropic trap with anharmonic energy levels. If 
one then assumes that the resulting energy level shifts are 
larger than cooling rates, one can evoke standard secular 
arguments to reduce the master equation to a diagonal form in 
the basis of the {\it bare} ideal gas states \cite{CLZ}. It is 
precisely  the subject of the present paper to study the 
situation in which such reduction is not possible, i.e. when 
the effects of accidental degeneracy dominate the dynamics of 
the system. 

One could argue that such problem is purely academic since in real
physical system atom-atom interaction 
will always lift the accidental degeneracy.
The whole point is, however, that as long as the interactions are not
too strong, the system will still exhibit effects of {\it quasi-degenracy}.
This will be the case when energy differences between the quasi-degenerated levels will be small in comparison to $\hbar\omega$. Such condition is realised
in not too dense systems, i.e. the system containing not too many atoms.
The idealized theory presented in this paper is formulated
for arbitrary number of atoms $N$, but in practice it 
applies to the  situation when $N$ is not too large (see Section  VII). 
Nevertheless, in view of the complexity of the problem, it is in our opinion
reasonable to treat the ideal case 
of non-interacting atoms in order to get insight into more realistic cases.
That said, let us note at this point that in the last years
the external modification of the 
s--wave scattering length (which dominates 
the atom--atom collisions at low energies), 
has been theoretically investigated in different scenarios \cite{asctheo}, 
and also experimentaly demonstrated by employing the so--called 
Feshbach resonances \cite{ascMIT,ascJILA}.
Remarkably, very recent experimental results show that a modification of the 
scattering length to very small values is experimentally feasible 
\cite{ascJILA}, opening the fascinating possibility to acheive  
a quasi--ideal bosonic gas, as that studied in the 
present paper.

\subsection{Content of the paper}

The paper concerns thus the problem of cooling of an {\it ideal} Bose gas,
or more precisely speaking a sample consisting of $N$ atoms, 
in a {\it perfectly harmonic microtrap}. It should be noted that  
this problem  is quite general. Atomic traps, although 
frequently anisotropic (see for example Ref. \cite{Eric}), can 
be designed to be, with a very good accuracy, harmonic. Moreover, even though 
in the small
atomic samples atom-atom interactions will lift the exact degeneracy
of energy levels, the system will remain {\it quasi-degenerated}. We 
expect that a cooled atomic sample in such a harmonic microtrap 
will exhibit the effects of accidental quasi-degeneracy {\it 
regardless of the method used for its cooling}
 In order to 
stress this general aspect of our study, we adopt here 
partially the statistical physics approach, and  derive a 
master equation using a phenomenological model of the ``bath''. 
In particular, our ``bath'' may  represent one of the following two 
reservoirs: B1) a photon reservoir in the case 
of laser cooling \cite{CLZ}; B2) an atomic reservoir consisting of other 
atoms in the case of symphatetic cooling \cite{sympa}. In the  case  B2) 
the bath atoms are trapped in a larger trap than the system atoms. 
 Such  situation  (proposed and discussed in \cite{sympa})
 could be realised if a small, say far-off-resonance dipole  trap for system 
atoms was located inside a larger magnetic trap for the bath atoms 
\cite{garzol}. 

Given that the 
resulting master equations for all these reservoirs have the 
same structure, the qualitative behavior for the cooling 
dynamics given by our specific model is quite general. The 
reason why we have chosen such a model for the  bath is that it 
has the adventage that one can derive analytical formulas for 
the transition rates between different levels. We also stress 
that the mathematical treatment developed in this paper can be 
easily generalized to other physical situations, in particular for 
ultra--cold trapped polarised Fermi gases, which due to the suppression 
of the $s$--wave scattering induced by the Pauli principle, can be 
in an excelent approximation as ideal gases \cite{Jin}.

This paper is organized as follows: In Section II we introduce 
the model, describing separately  the system in a trap, the 
atomic bath, and the system--bath interactions. This Section is 
very much analogous to Section II of Ref. \cite{sympa}, but is 
formulated for a different model of the bath.  In Section III 
we derive the master equation governing the dynamics of the 
system,  under Born and Markov approximations. This equation is 
further analyzed  in Section IV in the microtrap limit in terms 
of the Lamb-Dicke (LD) expansion, i.e. a systematic  expansion 
in a small parameter $\eta=ka$, where $a$ is the size of the 
trap ground state wavefunction, whereas $k$ is a typical 
wavevector of the bath quanta that is relevant for the cooling 
process. In Section V we discuss explicitely the breakdown of 
ergodicity due to the accidental degeneracy that occurs on a 
fast time scale in the lowest order of the LD expansion. 
Section VI is devoted to the discussion of the restoration of 
ergodicity on a slower time scale due to   higher orders of LD 
expansion, whereas Section VII contains our conclusions. The 
paper also contains three appendices. Appendix A contains 
some useful formulas of the operator algebra used in the paper,
and describes the structure of the Hilbert space.
Appendix B describes the 
details of the construction of some of the
 multiple vacua that appear due 
to the accidental deneracy, whereas Appendix C presents 
matrix elements of the second quantized operators used in 
the paper. These elements are used for some calculations  
regarding the cooling rates. Finally, in Appendix D we present 
a proof of the decay of coherences on the slow time scale.

% -----------------------------------------------------------
\section{Description of the model}
% -----------------------------------------------------------

We consider a system ``A" of bosonic particles  that are 
confined in a trap, and interact with a bath  bosonic 
particles  ``B". We assume that the particles ``B"  are 
practically not affected by their interactions with the system 
``A'', so that ``B'' can be regarded as a phenomenological 
reservoir for ``A''. The coupling to the bath represents the 
influence of some externally controlled cooling mechanism  
(laser cooling, symphatetic cooling, etc.) 
on the system ``A''. The reservoir is assumed  to be  in 
thermal equilibrium at some given temperature.  In this Section 
we introduce the Hamiltonian for the system ``A", the bath 
``B",  and for their mutual interactions. The formalism is 
developed for the case of $d$ spatial dimensions.

% -----------------------------------------------------------

\subsection{Description of the system}

The system ``A" is an {\it ideal gas} of $N$ bosonic atoms of 
mass $M_A$ confined by an isotropic harmonic potential in $d$ 
dimensions. In a second quantized form, and in the Fock 
representation  the Hamiltonian describing the system can be 
written in the form
\begin{equation}
\label{Habla}
H_A = \sum_{\vec n} \hbar \omega (n_x+n_y+\ldots) 
a^\dagger_{\vec n} a_{\vec n},
\end{equation}
where $\omega$ is the trap frequency, ${\vec 
n}=(n_x,n_y,\ldots)$ with $n_x,n_y,\ldots = 0,1,2,\ldots$, and 
$a^\dagger_{\vec n}$ and $a_{\vec n}$ are creation and 
annihilation operators of particles in the ${\vec n}$--th level 
of the harmonic potential, respectively. Since in the present 
paper we are interested in the effects of the accidental 
degeneracy, we neglect the contribution of the atom--atom 
interactions to the total Hamiltonian (for the discussion of 
its role see \cite{CLZ,gbed,sympa}, and Section VII).

\subsection{Bath}

Similarly as in Ref. \cite{sympa}, we  assume that the system 
``B" contains  a practically  infinite number of bosons of mass 
$M_B$ embedded in a practically infinite volume, with finite 
density $n_B$. The free Hamiltonian for the bath ``B" of 
particles in a second quantized form is
\begin{equation}
H_B = \int d {\vec k} \epsilon({\vec k}) 
b({\vec k})^\dagger b({\vec k}).
\label{eq1}
\end{equation}
Here, ${\vec k}$ is a wavevector in a $d$--dimensional space. 
The function $\epsilon(\vec k)$ represents the dispersion 
relation of  the bath particles. For instance, for quasi-free 
particles it reads
\begin{equation}
\label{Ebath}
\epsilon ({\vec k}) = \frac{(\hbar { k})^2}{2 M_B},\label{2ed}
\end{equation}
where $M_B$ is the {\it effective mass} of the bath ``quanta''. 
In the following we shall use Eq. (\ref{2ed}) but the theory 
is easily generalized to arbitrary shapes of the dispersion 
relation. In particular, we shall also  consider massless bath quanta 
with a photonic-like linear dispersion relation. Note that such dispersion relation is in fact appropriate for both types of bath (B1), and (B2) mentioned above. This is obvious in the case of laser cooling, but is also true for collisional cooling schemes, since no mass is produced or lost in the collision processes.   The operators 
$b({\vec k})^\dagger$ and $b({\vec k})$ are creation and 
annihilation operators of bath particles corresponding to  
plane wave states with momentum ${\vec k}$. In the case of laser cooling (B1)
${\vec k}$ corresponds to  a photon momentum, whereas in the case (B2)  ${\vec k}$ 
is rather  a {\it momentum transfer} associated with the system atom-bath atom collision.
The operators
$b({\vec k})^\dagger$ and $b({\vec k})$  fulfill the 
usual commutation relations
\begin{eqnarray}
\left[b({\vec k}),b({\vec k}')\right] &=& 
\left[b({\vec k})^\dagger,b({\vec k}')^\dagger\right]=0,\nonumber\\
\left[b({\vec k}),b({\vec k}')^\dagger\right] &=& 
\delta^{(d)}({\vec k}-{\vec 
k}').\nonumber
\end{eqnarray}

In thermal equilibrium, the density operator describing the 
state of the bath $\rho_B$ corresponds to the usual Bose--Einstein 
distribution (BED) \cite{Groot}. In this situation, we have
\begin{mathletters}
\label{bbdag}
\begin{eqnarray}
\langle b({\vec k}) b({\vec k}') \rangle &=&                
\langle b({\vec k})^\dagger b({\vec k}')^\dagger \rangle = 0,\\
\langle b({\vec k})^\dagger b({\vec k}') \rangle &=& 
n({\vec k}) \delta^{(d)} ({\vec k}-{\vec k}'),
\end{eqnarray}
\end{mathletters}
where $n({\vec k})$ is related to the number of particles 
with wavevector ${\vec k}$, and is given by
\begin{equation}
n({\vec k})=\frac{z e^{-\beta\epsilon({\vec k})}}
{1-z e^{-\beta \epsilon({\vec k})}}.
\end{equation}
In the above expression, $\beta=1/(k_B T)$ is the 
inverse temperature, and $z=\exp(\beta\mu)$ is the fugacity, 
while $\mu$ denotes the chemical potential. Note that $\mu=0$, 
$z=1$ for  massless quanta, and for  massive particles below 
the condensation point. Note also that both $n({\vec k})$ and 
$\epsilon({\vec k})$ only depend on $k\equiv |{\vec k}|$. 
Particle and energy densities are connected with these 
quantities by the relations
\begin{mathletters}
\begin{eqnarray}
n_B &=& \frac{1}{(2\pi)^d} \int d{\vec k} n({\vec k}),\\
\epsilon_B &=& \frac{1}{(2\pi)^d} \int d{\vec k} n({\vec k}) 
\epsilon({\vec k}),
\end{eqnarray}
\end{mathletters}
respectively.

\subsection{Interactions}

Within the present model the interactions between the particles 
and the bath describe the annihilation of an atom followed by 
its inmediate recreation, accompanied by absorption or emission 
of a single bath quantum. We have chosen such a model for the 
interactions since it is the simplest one that retains the effect 
produced by the accidental degeneracy in the cooling process. 
In any case, it may be regarded as a phenomenological 
interaction that may be due to various physical mechanisms 
(system atom-bath atom collisions, 
laser-atom interactions, etc.). Similarly to  the case of 
atom-atom collisions we employ here an analogue  of the {\it 
shape--independent approximation} \cite{Huang,many} to write 
down the corresponding atom-bath quantum interactions. We 
assume that these interactions are local (i.e. have a zero 
range) in the spatial representation. Mathematically, this 
approximation means that the wave functions of both kinds of 
particles do not change significantly over distances 
characterizing interparticle potential in the relevant energy 
range.  In the Fock representation the interaction Hamiltonian 
is given by 
\begin{equation}
\label{HAB}
H_{A-B} = \sum_{{\vec n},{\vec n}'} \int d{\vec k} 
\gamma_{{\vec n},{\vec n}'}({\vec k}) 
a_{\vec n}^\dagger a_{{\vec n}'} b^\dagger({\vec k}),
\end{equation}
where 
\begin{equation}
\gamma_{{\vec n},{\vec n}'}({\vec k}) =\frac{\kappa}{(2\pi)^d}
\int d{\vec x} \psi_{\vec n}({\vec x})^\ast \psi_{{\vec n}'}({\vec x}) 
e^{-i{\vec k}{\vec x}},
\end{equation}
$\psi_{\vec n}({\vec x})$ is the wavefunction corresponding to 
the ${\vec n}$--th level of the harmonic oscillator, and $\kappa$ is 
a coupling constant. We have chosen the form of the interaction 
Hamiltonian (\ref{HAB}), because of its simplicity. We stress, 
however, that the qualitative (and to some extend quantitative) 
results of the paper do not depend on the particular choice of 
$H_{A-B}$.

Without  loss of generality, we can exclude from the integration 
over ${\vec k}$  in (\ref{HAB}) the value 
${\vec k}=0$. This is clear since
\begin{eqnarray}
&&\sum_{{\vec n},{\vec n}'} 
\gamma_{{\vec n},{\vec n}'}(0) 
a_{\vec n}^\dagger a_{{\vec n}'} b^\dagger({0}) \propto \nonumber \\
\label{constant}
&& \left(\sum_{{\vec n}} 
a_{\vec n}^\dagger a_{{\vec n}}\right)\left(
b^\dagger({0}) \right),
\end{eqnarray}
is a constant shift  operator of the zero momentum mode, 
proportional to the number of particles in the system ``A''. 
One can always perform a unitary shift transformation of 
$b(0)$, and $b^{\dag}(0)$ that cancels the term 
(\ref{constant}) and its hermitian conjugate   in the 
Hamiltonian. Obviously, such transformation modifies the BED 
for the bath quanta with zero momentum, but the latter 
modification has no relevance for the transformed system-bath 
interactions in which the coupling to the bath zero mode is 
absent. Hence, from now on, in the integrals over ${\vec k}$ it 
will be implicitely assumed that ${\vec k}\ne 0$. On the other 
hand, since in the next Section we are going  to make a 
rotating wave approximation (RWA) in the master equation 
derived from the  Hamiltonian (\ref{HAB}), we reduce  Eq. 
(\ref{HAB}) (as in Ref. \cite{sympa}) to the form 
\begin{equation}
\label{hrwa}
H_{A-B} = H_0 + \sum_{\alpha=1}^\infty (H_\alpha + 
H_\alpha^\dagger).
\end{equation}
Here, $H_0$ contains the part of $H_{A-B}$ given in (\ref{HAB}) 
in which the sum is extended over values with 
$\sum_{s=x,y,\ldots} (n_s-n'_s)=0$. $H_\alpha$ contains the 
part of $H_{A-B}$ proportional to the bath annihilation 
operators $b({\vec k})$, with the sum  over $\vec n$, ${\vec 
n}'$ extended over the values for which $\sum_{s=x,y,\ldots} 
(n_s-n'_s)=\alpha$.  

The QME that we shall derive in the next Section 
will contain a Hamiltonian part (describing energy level 
shifts), and non-Hamiltonian part describing dissipative decay 
processes.  It is worth mentioning that due to the RWA the 
Hamiltonian part of the master equation will not be generally 
correct \cite{Gardinerbook,Cohen}. However, as in Ref. 
\cite{sympa}, we will be only interested in the dissipative 
part of the master equation, which is correctly described  
under the mentioned RWA, provided the trap frequency $\omega$ 
is larger than the cooling rates. The latter  assumption  will 
be made all over the present paper.

\section{Derivation of the master equation}

The master equation for the above defined model can be derived 
following well--stablished procedures in the field of quantum 
optics \cite{Gardinerbook,Cohen}, analogously to those 
discussed in Ref. \cite{sympa}. We first move to an interaction 
picture defined by the unitary operator $\exp{[-i (H_A + 
H_B)t]}$. In this picture, the density operator $\tilde\rho$ 
describing system--plus--bath degrees of freeedom fulfills the 
following equation:
\begin{equation}
\label{VNEq}
\frac{d\tilde\rho(t)}{dt} = -\frac{i}{\hbar}[\tilde 
H_{A-B}(t),\tilde \rho(t)],
\end{equation}
where the tilde indicates that the operators are expressed in 
the interaction picture. We integrate formally this equation, 
and substitute back into (\ref{VNEq}). Next, we define the 
reduced density operator for system ``A", $\rho_A={\rm Tr}_B 
(\rho)$, where Tr$_B$ stands for the trace over the bath 
states,  and make use of the fact that 
Tr$_B\{[H_{A-B}(0),\rho(0)]\}=0$, since we assume that the 
density operator for the bath $\rho_B(0)$ is diagonal in the 
Fock basis (with respect to $H_B$), whereas $H_{A-B}$ does not 
contain any diagonal matrix elements [the reader should recall 
that we have extracted the terms with ${\vec k}=0$ in 
$H_{A-B}$, see Eq. (\ref{HAB})]. We obtain the following 
equation:
\begin{eqnarray}
&&\frac{d{\tilde\rho_A(t)}}{dt} =
\nonumber \\ 
&&-\frac{1}{\hbar^2}\int_0^t d\tau
{\rm Tr}_B \{[\tilde H_{A-B}(t),[\tilde H_{A-B}(t\!-\!\tau),\tilde 
\rho(t\!-\!\tau)]] \}.
\label{VNEq3}
\end{eqnarray}
\noindent

In the following step we perform Born--Markov 
approximation. For this approximation we have to assume 
that the correlation time $\tau_c$ of the reservoir is much 
shorter than the typical time over which $\tilde\rho_A(t)$ 
changes, i.e. the cooling time \cite{Cohen}. The Born--Markov   
approximation is also related to the fact that the bath quanta are 
practically not affected by their interactions with the system; 
this allows us to write $\tilde\rho(t-\tau)= 
\tilde\rho_A(t-\tau)\otimes\rho_B(0)$.  From the technical 
point of view, the correlation time $\tau_c$ can be defined as 
a time for which  the integrand of (\ref{VNEq3}) practically 
vanishes. For specific physical models of the bath it can be 
directly evaluated  (see, for instance, Ref. \cite{sympa}).
The cooling  time, on the other hand,  depends on the physics 
of the interactions  between atoms and bath quanta. It is {\it 
controllable}, and thus  is assumed to be the longest time 
scale of the problem.   In this case, we can safely substitute 
$\tilde\rho_A(t-\tau)$ by $\tilde\rho_A(t)$ in the integral 
(\ref{VNEq3}), and extend the upper limit of the integral to 
infinity. 

In the next steps we make  use of Eq. (\ref{hrwa}) and  perform 
the RWA, i.e. neglect terms rotating at multiples of the trap 
frequency.  Again, this approximation is based on the 
assumption that trap frequencies are large in comparison to the 
cooling rates.

Finally, taking into account the bath properties (\ref{bbdag}), 
and coming back to the Schr\"odinger  picture we obtain the 
following master equation:
\begin{equation}
\label{ME}
\frac{d\rho_A}{dt} = -\frac{i}{\hbar}[H_A +H_{A-A}',\rho_A]
+ {\cal L}\rho_A,
\end{equation}
where $H_{A-A}'$ is  the Hamiltonian term produced by the 
elimination of the bath in the master equation.  Physically, 
this term   accounts for the energy level shifts due to the 
effective interaction between system particles via their 
interactions with  the bath quanta. $H_{A-A}'$ may also include 
the original atom-atom interactions (provided they were present 
in the original Hamiltonian) in the spirit of {\it independent 
rates approximation} \cite{Cohen}.  Similarly as in the case of 
the atom--atom collisions, all of the shifts caused by 
$H_{A-A}'$ may be neglected in some situations depending on the 
specyfic model of the bath, the size of the trap, and the 
number of atoms in the system (for details see 
\cite{CLZ,gbed,sympa}). We shall omit them in the following, 
and come back to the discussion of their role in Section VII. 
We shall therefore postulate a QME restricting our attention to 
the Liouvillian ${\cal L}$ that describes the cooling process,
\begin{equation}
{\cal L} = \sum_{\alpha\ne 0}^\infty {\cal L}_\alpha,
\label{eq15}
\end{equation}
where
%\widetext
\begin{eqnarray}
\label{Lalpha}
{\cal L}_\alpha \rho_A &=& \sum_{{\vec n},{\vec n}',{\vec m},{\vec m}'} 
\Gamma_{{\vec n},{\vec n}'}^{{\vec m},{\vec m}'} 
(2a^\dagger_{\vec m} a_{{\vec m}'} \rho_A a^\dagger_{\vec n} a_{{\vec n}'}
\nonumber \\
&-& a^\dagger_{\vec n} a_{{\vec n}'} a^\dagger_{\vec m} a_{{\vec m}'} \rho_A
- \rho_A a^\dagger_{\vec n} a_{{\vec n}'} a^\dagger_{\vec m} a_{{\vec m}'}).
\end{eqnarray}
%\narrowtext
The sum in this expression is extended to ${\vec n},{\vec n}',
{\vec m},{\vec m}'$ fulfilling
\begin{equation}
\label{alpha}
\sum_{s=x,y\ldots} (n_s-n'_s)=\alpha, \;\;\;\; 
\sum_{s=x,y\ldots} (m_s-m'_s)=-\alpha.
\end{equation}
Liouvillian (\ref{Lalpha}) acounts for transitions of  
particles from one level of the harmonic oscillator to another, 
experiencing a change in the energy of $\alpha\hbar\omega$, and 
a corresponding absorption or emission of a bath quantum. Thus, 
the term with $\alpha=0$ that conserves the energy does not 
enter (\ref{eq15}) since we have excluded the bath quanta with 
zero energy from the interaction (see Sec. II). The terms with 
$\alpha>0$ ($\alpha<0$), on the other hand, describe processes 
increasing (decreasing) the energy. These transitions are 
characterized by 
\begin{eqnarray}
\label{crates}
\Gamma_{{\vec n},{\vec n}'}^{{\vec m},{\vec m}'} &=&
\frac{\pi}{\hbar} \int d{\vec k}  
\gamma_{{\vec n},{\vec n}'}({\vec k}) 
\gamma^*_{{\vec m},{\vec m}'}({\vec k})
\nonumber\\
&&\times [n({\vec k})+1]
\delta[\epsilon(k)-\hbar\omega\alpha],
\end{eqnarray}
for  $\alpha$ positive, and by
\begin{eqnarray}
\label{crates2}
\Gamma_{{\vec n},{\vec n}'}^{{\vec m},{\vec m}'} &=&
\frac{\pi}{\hbar} \int d{\vec k}  
\gamma_{{\vec n},{\vec n}'}({\vec k}) 
\gamma^*_{{\vec m},{\vec m}'}({\vec k})
\nonumber\\
&&\times n({\vec k})
\delta[\epsilon(k)-\hbar\omega|\alpha|],
\end{eqnarray}
for $\alpha$ negative.

We stress here the fact that for any kind of interactions 
between the atoms and the bath, the corresponding Liouvillian 
has the same form as in (\ref{eq15}) and (\ref{Lalpha}), 
provided in each interaction act one atom is annihilated and 
another created. This is the reason why our results of the 
following sections can be extended to other kind of 
interactions. Note, however, that in such cases the 
coefficients (\ref{crates}) and (\ref{crates2}) may have a much 
more complicated form (see, for example, Ref. \cite{sympa}).

\section{Lamb-Dicke expansion  in 1D}

In this Section we perform the Lamb-Dicke (LD) expansion of the 
master equation (\ref{eq15}). From now on we shall concentrate 
on the one dimensional case and skip the vector notation.

The LD expansion is valid in the situation when the bath quanta 
relevant for the cooling process have momenta $\vec k$ much 
smaller than the inverse of the size of the trap, $a$. Their 
corresponding wave functions [$\propto \exp(\pm i k x)$] vary 
slowly on a scale of $a$, and can be then expanded in  
Taylor series around $x\simeq 0$. Since, according to Eqs. 
(\ref{crates}) and (\ref{crates2}) the relevant bath quanta 
have energies $\epsilon(k)=\hbar\omega\alpha$, their 
corresponding momenta may be determined from the dispersion 
relation. For example, for the case of  massive free particles 
characterized by the dispersion relation (\ref{Ebath}), the 
validity of LD at relatively low temperatures requires that
\begin{equation}
\left(\frac{2M_B\omega|\alpha|}{\hbar}\right)^{1/2}\left(
\frac{\hbar}{2 M_A\omega}\right)^{1/2}=
\left(\frac{M_B|\alpha|}{M_A}\right)^{1/2}
< 1,
\label{conld}
\end{equation}
i.e. the effective mass of the bath particles must be much 
smaller than that of the system atoms \cite{noteii}. It is easy to find 
analogous conditions corresponding to other forms of the 
dispertion relation, including the case of massless bath quanta.
In case of the laser cooling (B1) at low temperatures
the condition is simply
that $ka=2\pi a/\lambda<1$, where $k$ ($\lambda$) is the laser photon momentum 
(wavelength). 

In the case of symphatetic cooling (B2)
the condition is
$\Delta k a <1$, where $\Delta k$ is the typical momentum transfer
in a collision act. In such collision the kinetic
energy of the incoming bath atom is of the order of $1/\beta$, and 
its correponding momentum  of the order of $2\pi/\lambda_B$, where $\lambda_B$
denotes the  thermal de Broglie wavelength. If the bath was cool, $\beta\hbar\omega \gg 1$, the momentum transfer would typically be equal to the final momentum of the bath atom. The validity of LD expansion would then require
that the same condition as (\ref{conld}) is fulfilled, with $M_B$ denoting the real mass of the bath atoms, and that $2\pi a/\lambda_B<(M_B/M_A)^{1/2}<1$.

In the LD regime we can expand the rates (\ref{crates}), 
(\ref{crates2}). Denoting by
\begin{eqnarray}
k_{\alpha}&&=\left(\frac{2M_B\omega|\alpha|}{\hbar}\right)^{1/2}, \\
\eta_{\alpha}&&=k_{\alpha}a\equiv\sqrt{\alpha}\eta,
\label{deno}
\end{eqnarray}
we obtain for $n$, $n'$, $m$, $m'$ fulfilling Eq. (\ref{alpha})
\begin{eqnarray}
\Gamma^{m,m'}_{n,n'}&=&\frac{2\pi}{\hbar} \gamma_{n,n'}(k_{\alpha})
\gamma^*_{m,m'}(k_{\alpha})
\nonumber \\
&\times&[n(k_{\alpha})+
(1+{\rm sign}(\alpha))/2]\frac{M_B}{\hbar^2k_{\alpha}},
\end{eqnarray}
and
\begin{eqnarray}
\gamma_{n,n'}(k_{\alpha})&=&\frac{\kappa}{2\pi}\Bigl[\delta_{n,n'}
-\! i\eta_{\alpha}\left(\sqrt{n+1}\delta_{n,n'-1}+
\sqrt{n}\delta_{n,n'+1}\right)
\nonumber \\
&-& \left. \! \frac{\eta_{\alpha}^2}{2} \! \left(\sqrt{(n+2)(n+1)}
\delta_{n,n'-2} +
(2n+1)\delta_{n,n'}\right.\right.
\nonumber \\
&+&\left.\left.\sqrt{n(n-1)}
\delta_{n,n'+2}\right) +\ldots \right]
\label{expld}
\end{eqnarray}
The master equation can be now rewritten using the above 
expressions in the interaction picture with respect to $H_A$, 
and upon neglection of $H'_{A-A}$.  It takes then the following 
form
\begin{equation}
\dot\rho=\left[{\cal L}^{(0)}+{\cal L}_1^{(1)}+
{\cal L}_2^{(1)}+ O(\propto \eta^6)+ \ \ldots\right]\rho.
\label{masld}
\end{equation}

The ``zeroth'' order term is actually of the order of $\eta^2$ 
and has the form
\begin{eqnarray}
{\cal L}^{(0)}\rho&&= \Gamma_{\downarrow}\left[2A\rho A^{\dag}
-A^{\dag}A\rho - \rho A^{\dag}A\right] \nonumber\\
&&+  \Gamma_{\uparrow}\left[2A^{\dag}\rho A
-AA^{\dag}\rho - \rho A A^{\dag}\right];
\label{lzero}
\end{eqnarray}
where
\begin{mathletters}
\begin{eqnarray}
\Gamma_{\downarrow}&=&N\frac{\kappa^2}{(2\pi)^2}
\frac{2\pi}{\hbar}\eta^2\left[n(k_1)+1
\right]\left(\frac{M_B}{\hbar^2k_1}\right), \label{gammad}\\
\Gamma_{\uparrow}&=&N\frac{\kappa^2}{(2\pi)^2}
\frac{2\pi}{\hbar}\eta^2\left[n(k_1)
\right]\left(\frac{M_B}{\hbar^2k_1}\right), \label{gammau}
\end{eqnarray}
\end{mathletters}
whereas
\begin{mathletters}
\begin{eqnarray}
A&&=\frac{1}{\sqrt{N}}\sum_{n=0}^{\infty}\sqrt{n+1}a^{\dag}_na_{n+1},
\label{defa} \\
A^{\dag}&&= \frac{1}{\sqrt{N}}\sum_{n=0}^{\infty}
\sqrt{n+1}a^{\dag}_{n+1}a_{n}.
\label{defad}
\end{eqnarray}
\end{mathletters}

Note that quite generally: a) both cooling rates 
$\Gamma_{\downarrow}$, $\Gamma_{\uparrow}$ are collective (i.e. 
proportional to $N$); b) their ratio is
\begin{equation}
\frac{\Gamma_{\uparrow}}{\Gamma_{\downarrow}}=\frac{n(k_1)}{n(k_1)+1}=
z e^{-\beta\hbar\omega}. \label{ratiog}
\end{equation}

Similarly, the higher order terms (of order $\propto \eta^4$) are
\begin{eqnarray}
{\cal L}^{(1)}_1\rho&&= 
-\Gamma_{\downarrow}\eta^2\left[2\left(A\rho C^{\dag}
+ C\rho A^{\dag}\right)\right.
\nonumber \\
&-&\left.\left(C^{\dag}A\rho + A^{\dag}C\rho\right) 
- \left(\rho C^{\dag}A + \rho A^{\dag}C\right)\right] \nonumber\\
&-&\Gamma_{\uparrow}\eta^2\left[2\left(A^{\dag}\rho C
+ C^{\dag}\rho A\right)\right.
\nonumber \\
&-&\left.\left(CA^{\dag}\rho + AC^{\dag}\rho\right) 
- \left(\rho CA^{\dag} + \rho AC^{\dag}\right)\right],
\label{loneone}
\end{eqnarray}
with
\begin{mathletters}
\begin{eqnarray}
C=&&\frac{1}{2\sqrt{N}}\sum_{n=0}(n+1)^{3/2}a^{\dag}_{n}a_{n+1}, \\
C^{\dag}=&&\frac{1}{2\sqrt{N}}\sum_{n=0}(n+1)^{3/2}
a^{\dag}_{n+1}a_n \label{dupkl}, 
\end{eqnarray}
\end{mathletters}
and 
\begin{eqnarray}
{\cal L}^{(1)}_2\rho&&= \Gamma^{(1)}_{\downarrow}\left[2B\rho B^{\dag}
-B^{\dag}B\rho - \rho B^{\dag}B\right] \nonumber\\
&&+  \Gamma^{(1)}_{\uparrow}\left[2B^{\dag}\rho B
-BB^{\dag}\rho - \rho BB^{\dag}\right];
\label{lonetwo}
\end{eqnarray}
where
\begin{mathletters}
\begin{eqnarray}
\Gamma^{(1)}_{\downarrow}=N\frac{\kappa^2}{(2\pi)^2}
\frac{2\pi}{\hbar}\frac{\eta_2^4}{4}
\left[n(k_2)+1
\right]\left(\frac{M_B}{\hbar^2k_2}\right), \label{gammado}\\
\Gamma^{(1)}_{\uparrow}=N\frac{\kappa^2}{(2\pi)^2}
\frac{2\pi}
{\hbar}\frac{\eta_2^4}{4}\left[n(k_2)
\right]\left(\frac{M_B}{\hbar^2k_2}\right), \label{gammauo}
\end{eqnarray}
\end{mathletters}
whereas
\begin{mathletters}
\begin{eqnarray}
B&&=\frac{1}{\sqrt{N}}\sum_{n=0}^{\infty}
\sqrt{(n+1)(n+2)}a^{\dag}_na_{n+2},
\label{defao} \\
B^{\dag}&&= \frac{1}{\sqrt{N}}\sum_{n=0}^{\infty}
\sqrt{(n+1)(n+2)}a^{\dag}_{n+2}a_{n}.
\label{defado}
\end{eqnarray}
\end{mathletters}
Note that similarly as in Eq. (\ref{ratiog}) the rates fulfill
\begin{equation}
\frac{\Gamma^{(1)}_{\uparrow}}{\Gamma^{(1)}_{\downarrow}}=
\frac{n(k_2)}{n(k_2)+1}=
z e^{-2\beta\hbar\omega}. \label{ratiogt}
\end{equation}
Finally, Eq. (\ref{masld}) contains terms of higher orders 
$\propto \eta^6,\eta^8$, etc.

\section{Accidental degeneracy and the breakdown of ergodicity}

In this Section we discuss the dynamics governed 
by the lowest order term in the Lamb-Dicke expansion, i.e. by the equation
\begin{equation}
\dot \rho = {\cal L}^{(0)}\rho.
\label{dynzer}
\end{equation}

First we observe that
\begin{equation}
[A, A^{\dag}]=1,
\label
{harmo}
\end{equation}
i.e. the operators $A$, and $A^{\dag}$ represent an abstract 
harmonic oscillator Weyl--Heisenberg algebra, whereas the 
Liouville-von Neuman superoperator (\ref{lzero}) describes {\it interaction 
of that harmonic oscillator with an effective  thermal bath} 
(see, for instance \cite{Gardinerbook}). The inverse 
temperature of the bath is given by 
\begin{equation}
\hbar\omega\beta_{e}=
-\log\left(\frac{\Gamma_{\uparrow}}{\Gamma_{\downarrow}}\right)
=\hbar\omega\beta -\beta\mu.
 \label{ratiote}
\end{equation}
Note that the effective temperature is never greater than the 
temperature of the bath, $\beta_e\ge \beta$, since $\mu\le 0$. 
Alternatively, one may view the above equation, as if the 
effective temperature of the system was constant, but the 
frequency would change $\hbar\omega_e=\hbar\omega -\mu$. This 
effect is a result of our definition of the system--bath 
interactions. Those interactions consist of absorption or 
emission of the bath quanta, and thus do not conserve the 
number of the bath particles. The Boltzmann exponent 
(\ref{ratiote}) must account for energy gain or loss due to 
bath particles creation/annihilation.  The remaining question 
is how the abstract operator algebra is represented in the 
Fock-Hilbert space of our multiparticle system. To answer this 
question we first introduce the operator
\begin{equation}
D=B-A^2/\sqrt{N},
\end{equation}
which commutes with $A$, $A^{\dag}$, $B$ and $B^{\dag}$. We then observe
 that (see appendix A):
\begin{itemize}

\item For each  value of $l=0$, or $l=2,3, \ldots$ 
there exist $n_N(l)$, so called,  {\it vacuum} states $|0_{lsv}\rangle$ that 
are annihilated by the ``jump'' operator
$A=\sum_{n=0}^{\infty}\sqrt{n+1}g_n^{\dag} g_{n+1}$,
\begin{eqnarray}
A|0_{lsv}\rangle=0,
\end{eqnarray}
where the index $l$ indicates the {\it bare
energy} of the corresponding states (in units of $\hbar\omega$),
\begin{eqnarray}
\sum_{n=0}^{\infty} n a_n^{\dag}a_n|0_{l,s}\rangle= l |0_{l,s}\rangle,
\end{eqnarray}
The state corresponding to $l=0$ contains all $N$ atoms in the ground state.
The states with higher energy are constructed as linear 
combinations of degenerated Fock states. There is only one state of $l=1$
($N-1$ atoms in the ground state, and one atom in the first excited state),
which is not annihilated by $A$; that is why there is no vacuum state with
$l=1$.  
The index $s$ measures the number of excitations of $D^{\dag}$ in the state
$|0_{lsv}\rangle\propto (D^{\dag})^s|0_{l-2s0v}\rangle$, and runs from 0
 to $E(l/2)$, i.e. integer part of $l/2$. Each of the states $|0_{lsv}\rangle$
is an eigenstate of $D^{\dag}D$ with an eigenvalue
$(4ls-4s^2-2s +2Ns)/N$. The index $v$ enumerates finally the 
states of energy $l-2s$ annihilated by $D$. 
Denoting by $m_N(l)$ the number of states of energy
$l$ annihilated by $D$, i.e. the states 
\begin{equation}
D|0_{l0v}\rangle=0,
\end{equation}
we obtain
\begin{equation}
n_N(l)=\sum_{s=0}^{E(l/2)}m_N(l-2s).
\end{equation}
The  construction of the vacuum states is descibed in 
the Appendix A. Some other examples of the vacua are constructed in Appendix 
B. Each of the vacuum states is a 
linear combination of the {\it
accidentally degenerated} energy eigenstates. The number of 
accidentally degenerated states of the energy $ l$ in the $N$ atom
 system, $p_N(l)$,  is given by a solution of the partition problem
of the number theory \cite{hr18}
and  is extravagantly large
 (c.f. $p_N(l)\simeq O(\exp(\pi\sqrt{2l/3}))$
for $l\le N$). The number of the vacua is given by
$n_N(l)=p_N(l)-p_N(l-1)$. The very existence of
these multiple vacuum states is thus a direct consequence and, at the same
time, a signature of the {\it accidental degeneracy} \cite{flucties}.

\item The vacuum states are orthonormal, $\langle 0_{lsv}|0_{l's'v'}\rangle=
\delta_{ll'}\delta_{ss'}\delta_{vv'}$.

\item The Fock-Hilbert space of the system splits into an infinite number
of Fock 
subspaces corresponding to each of the vacuum states. The Fock states
in the $(l,s,v)$-th subspace are constructed as
\begin{eqnarray}
|k_{lsv}\rangle = \frac{(A^{\dag})^k}{\sqrt{k!}}|0_{lsv}\rangle,
\label{focki}
\end{eqnarray}
with $k=0,1,\ldots$. They are also mutually orthonormal, and are also
eigenstates of the energy operator with the corresponding eigenvalue
$(l+k)$. They are also highly degenerated (for $k+l=k'+l'$).
In the following we shall use the notation $w=(l,s,v)$.

\end{itemize}

The dynamics exhibits in the Lamb-Dicke limit multiple timescales. On the
fastest time scale of the order of $\eta^{-2}$ it is {\it nonergodic},
i.e. it does not mix the different $w$-subspaces.  After a short time
of the order of the inverse of 
$\Gamma_{\downarrow}$, $\Gamma_{\uparrow}$ 
 all coherences between the $|k_{w}\rangle$ and $|k'_{w'}\rangle$
for $k\ne k'$ vanish. Within each $w$-subspace the system approaches the
"thermal" equilibrium characterized by the density matrix  diagonal in
$k$, with zero off-diagonal elements, and 
the inverse  temperature $\beta_e$. The dynamics,
however, {\it cannot be reduced to a Poisson jump process}, i.e. a
sequence of random jumps between the various $|k_{w}\rangle$ states with
the transition probabilities governed by the detailed balance
conditions characteristic for the thermal equilibrium. The reason is
that coherences corresponding to $w\ne w'$, but $k=k'$ do not vanish.

The quantum mechanical density matrix in this (formally stationary)
limit becomes a sum of diagonal {\it canonical ensembles} in each of the subspaces, accompanied by a sum of non-diagonal terms connecting different
$w$ and $w'$ for the same $k$'s,
\begin{eqnarray}
&&\rho = \left[1-\exp(-\beta_{e}\hbar\omega)\right]\sum_{w}
\sum_k n_{w} |k_{w}\rangle\langle k_{{w}}|e^{-\beta_{e}\hbar\omega k} \nonumber\\
&&+\left[1-\exp(-\beta_{e}\hbar\omega)\right] \sum_{{w}\ne{w'} }
\sum_k r_{{w}{w'}}|k_{w}\rangle\langle k_{{w'}}|
e^{-\beta_e\hbar\omega k}.
\label{konie}
\end{eqnarray}
where the coefficients describe the populations of the corresponding 
subspaces, and are defined as
\begin{equation}
n_w= {\rm Tr}\left (P_w\rho(0)P_w\right),
\label{trace}
\end{equation}
and the cumulative coherences, 
\begin{equation}
r_{ww'}=\sum_k \langle k_w|\rho(0)| k_{w'}\rangle.
\label{fucin}
\end{equation}
Evidently, $\rho$ exhibits nonergodic effects and 
depends explicitely on the initial density operator. In the above expression
$P_w=\sum_k|k_w\rangle \langle k_w|$ denotes a projection operator onto the $w$-th subspace. Note that the definition (\ref{trace})
implies that $\sum_w n_w=1$.

We stress that in the Fock basis spanned by the states  
$|k_w\rangle$ the matrix $\rho$ is, in principle, not diagonal; 
moreover,  in general, depending on the initial condition, it 
does contain coherences when represented in the  Fock-Hilbert 
space corresponding to noninteracting atoms. That is the reason 
why the master equation in the latter basis cannot be reduced 
to the  diagonal form in the Lamb-Dicke limit. That is why in 
order to arrive at such reduction  additional assumptions have 
to be evoked that lift up the accidental degeneracy, as anharmonicity 
of energy levels caused by anharmonicity of the trap potential or
interatomic interactions.

\section{Restoration of ergodicity}

From the previous Section it is clear that the breakdown of 
ergodicity has only an approximate character since it is 
related to the lowest order dynamics in the LD expansion. It is 
natural to expect  that Eq. (\ref{konie}) describes only a 
quasi-stationatry solution which is indeed reached on a time 
scale $1/\Gamma_{\uparrow}, 1/\Gamma_{\downarrow} \simeq 
O(\propto 1/\eta^2)$, but undergoes further slow evolution on a 
time scale  $1/\Gamma^{(1)}_{\uparrow}, 
1/\Gamma^{(1)}_{\downarrow} \simeq O(\propto 1/\eta^4)$. On 
this slower time scale the ergodicity should be (at least 
partially) restored, and the transitions between the different 
$w$-th subspaces should become possible. As we shall see below, 
that is indeed the case.

To this aim we consider the higher order corrections to the 
master equation (\ref{masld}) and treat them perturbatively 
within the standard adiabatic elimination method \cite{Haake}. 
We introduce the projection operator
\begin{eqnarray}
&&{\cal P}\rho(t)=\sum_w n_w(t) \sum_k |k_w\rangle\langle k_w| e^{-
\beta_{e}\hbar\omega k}(1-  e^{-
\beta_{e}\hbar\omega})
\nonumber \\
&&+ \sum_{w\ne w'} r_{ww'}(t) \sum_k |k_w\rangle\langle k_{w'}| e^{-
\beta_{e}\hbar\omega k}(1-  e^{-
\beta_{e}\hbar\omega}), \label{duhyu1}
\end{eqnarray}
where the populations of the corresponding subspaces 
\begin{equation}
n_w(t)= {\rm Tr}\left(P_w\rho(t)P_w\right),
\label{nodt}
\end{equation}
and the cumulative coherences
\begin{equation}
r_{ww'}(t)=\sum_k \langle k_w|\rho(t)|k_{w'}\rangle,
\label{nodt1}
\end{equation}
become now slowly varying functions of time.

Introducing the complementary projector ${\cal Q}= 1 -{\cal P}$
obtain
\begin{eqnarray}
\dot {\cal P}\rho &&= {\cal P}\left({\cal L}_1^{(1)}+ {\cal L}_2^{(1)}
\right){\cal P}\rho + {\cal P}\left({\cal L}_1^{(1)}+ {\cal L}_2^{(1)}
\right){\cal Q}\rho , \label{piert} \\
\dot {\cal Q}\rho &&= {\cal Q}{\cal L}^{(0)}{\cal Q}\rho +
 {\cal Q}\left({\cal L}_1^{(1)}+ {\cal L}_2^{(1)}
\right){\cal P}\rho + \ldots \label{dotli}
\end{eqnarray}
In the latter Eq. (\ref{dotli}) we have already employed the 
fact that $\Gamma^{(1)}/\Gamma\simeq O(\propto \eta^2)$, and 
neglected the higher order terms in $\eta^2$. Moreover, 
adiabatic elimination of ${\cal Q}\rho$ from Eq. (\ref{dotli}) 
introduces corrections of the order $\propto\eta^6$ to Eq. 
(\ref{piert}) for ${\cal P}\rho$. Thus it may also be neglected 
in comparison to the leading term on the RHS of  Eq. 
(\ref{piert}).

We obtain thus a relatively simple master equation
\begin{equation}
\dot {\cal P}\rho = {\cal P}\left({\cal L}_1^{(1)}+ {\cal L}_2^{(1)}
\right){\cal P}\rho  \label{pierk} 
\end{equation}
From this master equation, after elementary algebra we obtain a 
set of closed rate equations for the populations $n_w$ of the 
$w$-th subspaces, and the coherences $r_{ww'}$. These equations 
can be enormously simplified using the properties of the 
operators $A$, $B$, $C$, and their hermitian conjugates that 
are discussed in Appendix A. Amazingly, it is only the term 
${\cal L}_2^{(1)}$ which contributes in this order of the LD 
expansion; moreover, the equations for populations and 
coherences decouple (see Appendix C for details),
\end{multicols}

\noindent \hrule width 8.6cm \vskip 4pt
\begin{eqnarray}
\dot n_w &=& 2\Gamma^{(1)}_{\downarrow} 
(1-e^{-\beta_{e}\hbar\omega}) \sum_{k} 
\left(\sum_{k',w'}|\langle k_w|B|k'_{w'}
\rangle|^2 e^{-\beta_{e}\hbar\omega k'}n_{w'}
- \langle k_w|B^{\dag}B|k_w\rangle  
e^{-\beta_{e}\hbar\omega k}n_{w}\right)
\nonumber\\
&+& 2\Gamma^{(1)}_{\uparrow}
(1-e^{-\beta_{e}\hbar\omega})\sum_{k}
\left( \sum_{k',w'}|\langle k_w
|B^{\dag}|k'_{w'}|^2 e^{-\beta_{e}\hbar\omega k'}n_{w'} -
\langle k_w|BB^{\dag}|k_{w}\rangle 
e^{-\beta_{e}\hbar\omega k}n_{w}\right),\\
\label{equar} 
\dot r_{ww'} &=& 2\Gamma^{(1)}_{\downarrow}
(1-  e^{-\beta_{e}\hbar\omega}) \sum_{k}
\left(\sum_{k',w_1,w_2} \langle 
k_w|B|k'_{w_1}\rangle \langle k'_{w_2}
|B^{\dag}|k_{w'}\rangle e^{-\beta_{e}\hbar\omega k'}r_{w_1w_2} 
\right.\nonumber \\
&-& \left. \frac{1}{2}\sum_{w_1}\left[\langle k_w
|B^{\dag}B|k_{w_1}\rangle r_{w_1w'}
+ \langle k_{w_1}|B^{\dag}B|k_{w'}\rangle r_{ww_1}  
\right] e^{-\beta_{e}\hbar\omega k}\right)
\nonumber \\
&+& 2\Gamma^{(1)}_{\uparrow} (1-  e^{-\beta_{e}\hbar\omega})
\sum_{k} \left( \sum_{k',w_1,w_2}
\langle k_w|B^{\dag}|k'_{w_1}\rangle 
\langle k'_{w_2}|B|k_{w'}\rangle 
e^{-\beta_{e}\hbar\omega k'}r_{w_1w_2}
\right.\nonumber \\
&-& \left. \frac{1}{2}\sum_{w_1}\left[\langle k_w
|BB^{\dag}|k_{w_1}\rangle r_{w_1w'}
+ \langle k_{w_1}|BB^{\dag}|k_{w'}\rangle r_{ww_1} \right]   
e^{-\beta_{e}\hbar\omega k}\right),
\label{equar1}
\end{eqnarray}

\pagebreak
\begin{multicols}{2}

The above equations are further reduced inserting the unity 
between $B$ and $B^{\dag}$, and using the fact that $A$, and 
$A^{\dag}$ by definition do not couple the different 
$w$-subspaces, whereas $B$ ($B^{\dag}$) has the only relevant  
matrix elements between the different $lsv$-subspaces for the 
 ($l\pm 2$, $s\pm 1$, $v$-th) and $lsv$-th subspace, 
for $l=2,3,\ldots$ 
($l=0,2,3,\ldots$), $s=0, \ldots, E(l/2)$, 
 and $k=k'=0,1, \ldots$ (see Appendix C).

From the above considerations we  obtain the final form of the
equations
\begin{eqnarray}
\dot n_w&=& 2\Gamma^{(1)}_{\downarrow}\left[f_{w+2}^2 n_{w+2}
- f_w^2 n_{w}\right] \nonumber \\
&+&  2\Gamma^{(1)}_{\uparrow}\left[f_w^2 n_{w-2}
- f_{w+2}^2 n_{w}\right].\label{glowna} \\
\dot r_{ww'}&=& 2\Gamma^{(1)}_{\downarrow}
\left[f_{w+2}f_{w'+2} r_{w+2w'+2}
- \frac{1}{2}\left(f_w^2\!+\!f_{w'}^2\right)
r_{ww'}\right] \nonumber \\
&+& 2\Gamma^{(1)}_{\uparrow}\left[f_wf_{w'} r_{w-2w'-2}
- \frac{1}{2}\left(f_{w+2}^2 \!+\!f_{w'+2}\right)
r_{ww'}\right], \nonumber \\
\label{glowna1}
\end{eqnarray}
where we have denoted $w=(lsv)$, $w\pm 2=(l\pm2,s\pm1,v)$,
with $f_w=\langle 0_{w-2}|B|0_w\rangle$. Note that $f_w$ can be 
assumed to be real and positive without loss of generality. As 
we see, both the populations $n_w$, and the coherences $r_{ww'}$  
fulfill a closed set of {\it rate equations}.  The explicit 
expressions for the matrix elements involved in the above 
formulae are derived in Appendix C.

The above equations provide the basic result of this paper. It 
shows that due to the presence of accidental degeneracy in the LDL the 
dynamics occurs essentially on several time scales. On the 
fastest time scale (governed by ${\cal L}^{(0)}$) the dynamics 
is nonergodic and consist in approach toward the canonical 
equilibrium states in each of the $l$-subspaces, accompanied by 
creation of quasi-equilibrium coherences between the states 
corresponding to the same $k$'s, but different $w$ and $w'$. 
On this scale the populations of each of the $w$-subspace, as 
well as the cumulative coherences $r_{ww'}$ may be regarded as 
constant, and therefore the values of these coherences, 
as well as the populations in 
each of the subspaces depend on the initial conditions.
In the higher order of expansion (on the time scale 
$1/\eta^2$ times longer), the mixing between different 
$w$-subspaces becomes possible. This mixing leads to a partial 
restoration of ergodicity. In the example considered here this 
restoration is not full, however, because as easily seen from 
Eq. (\ref{glowna}) the $w$-subspaces with different $v$ 
still do not mix. The reader can easily check that the further 
mixing of the odd and even subspaces does take place in the 
next order of the Lamb-Dicke expansion (for instance due to 
term containing bi-products of the operators $C$, and 
$C^{\dag}$).

The stationary state that is reached on the slower time scale 
is easy to find since Eqs. (\ref{glowna}) fulfill the detailed 
balance conditions, whereas the cumulative coherences are 
damped to zero, as demonstrated in Appendix D. We  obtain that
\begin{equation}
n_{l+2,s+1,v}= z e^{-2\beta\hbar\omega}n_{lsv},
\end{equation}
so that
\begin{eqnarray}
n_{2l+m,l,v}=e^{-2\beta'_{e}\hbar\omega l}n_{m0v},
\end{eqnarray}
with 
\begin{equation}
\label{iiig}
\beta'_e=\beta-\beta\mu/2\hbar\omega.
\end{equation}
The ratio of $n_{m0v}$, and $n_{m'0v'}$ remain undetermined in this 
order. Note that the reason why $\beta'_e\ne \beta_e$ is that both 
temperatures correspond to the processes that involve single 
bath quantum absorbtion or emission, but different energy 
changes (by $\hbar\omega$, or $2\hbar\omega$, respectively). 
Indeed, it is elementary to check that the stationary diagonal 
matrix elements of the desity matrix are proportional to the 
corresponding Boltzmann factors,
\begin{eqnarray}
&&\langle k_{2l+m,l,v}|\rho|k_{2l+m,l,v}\rangle \propto 
e^{-\beta_{e}\hbar\omega k -2\beta'_e\hbar\omega l}n_{m0v}\nonumber \\
&&= \langle k_{2l+m,l,v}|e^{-\beta'_{e}\hbar\omega 
\hat E-m+\beta\mu A^{\dag}A/2}n_{m0v}|k_{2l+m,l,v}\rangle .
\label{niewiem}
\end{eqnarray}

Finally, it is easy to check by substitution in the Liouvillian 
(\ref{eq15}) that for $\mu=0$ the steady state to all orders in the LD 
expansion is precisely $\rho \propto e^{-\beta H_A}$, which 
is diagonal in the original basis. 
Obviously, the steady state solutions obtained in the first, and the second 
order of our expansion are compatible with such a steady state.

% -------------------------------------------------------
\section{Conclusions}

In a series of papers \cite{CLZ,gbed,sympa,Dincol} we have studied in 
detail the quantum dynamics of bosonic and fermionic gases of 
cold atoms in  traps in the absence of accidental degeneracy. 
We studied various cooling mechanism, and various limiting 
cases. In this paper we have presented the solution of the 
corresponding problem accounting for accidental degeneracy 
effects. We have studied interactions of a gas of trapped atoms 
with a heat bath in the Lamb-Dicke limit using the master 
equation approach.  We have demonstrated  that the system 
approaches an equilibrium  on two (or more) distinguished 
time scales, and that the dynamics has the corresponding number 
of stages. At each stage a {\it quasi-equilibrium} state within 
appropriately determined subspaces of the Hilbert space is 
reached. This quasi-equilibrium corresponds to a canonical 
ensemble resticted to the appropriate subspace, and 
characterized by an effective temperature determined by the 
temperature and the chemical potential of the heat bath. In the 
next stage thermalization between the groups of subspaces 
occurs leading to another quasi-equilibrium in the larger 
subspaces, and so on.

We would like to stress once more that  the problem  considered 
in this paper is quite general. Atomic traps, although 
frequently anisotropic (see for example Ref. \cite{Eric}), can 
be designed to be harmonic with a very good accuracy. A cooled 
atomic sample  in such a harmonic microtrap will necessarily 
exhibit  the effects of accidental degeneracy {\it regardless 
of the method used for its cooling}!

One may question the generality of our results, since we have
used the Lamb-Dicke  expansion, and at the same time  
neglected in this paper atom-atom collisions, as well as 
atom-atom interactions mediated by the coupling with the bath. 
Such processes (described by the Hamiltonian $H_{A-A}'$, see 
Section III) lead evidently to shifts of the atomic energy 
levels, and will, in principle, lift up the accidental 
degeneracies. As we argued in Refs.  \cite{CLZ,gbed,sympa}, as 
long as the number of atoms in the microtrap is not too large, 
those shifts remain small and can be treated perturbatively. 
The system will then still exhibit the effects of accidental 
quasi-degeneracy. We stress that the theory developed in this 
article is valid for arbitrary numbers of atoms, 
and in particular it is for two atoms. Using far-off-resonance dipole traps
\cite{miller},
or loading atoms to a single minimum of a dark optical lattice (see Ref. \cite{boser}(b)) it should be accesible experimentaly to confine several atoms
in the trap of the size $a\simeq 0.1-0.05\mu$m. That implies validity of the LD expansion for the laser colling case (see \cite{CLZ}). Similarly, one can use
a cooled atomic gas close to, or below  the condensation point as a bath in the symphatetic cooling case. In the conditions of the experiments of Refs. \cite{Eric} and \cite{Bradley}
that implies de Broglie wavelength of the order of $\mu$m, and thus validates
the LD expansion. Using Bogoliubov-Hartree theory it is possible to estimate
perturbatively that the energy level splittings in a "band" of the quasi-degenerated states due to atom-atom collisions will be 
in such a case of the order of $N(a_{sc}/a)/\sqrt{N_D}$, where $a_{sc}$ is the scattering length of the system atoms, and $N_D$ is the number of levels in the band. Note that $N_D$ increases dramatically for higher excited levels.
We see with $a_{sc}=5$ nm, our theory should work for $N$ up to $\simeq 20$ even in the worst case when $N_D=2$. Note that the cooling of $20$ atoms to 
the ground state of a harmonic trap 
might be a very interesting task for the rapidly--developing research field of 
quantum informacion processing. Additionaly, we want to recall at this point 
that as pointed above, the external modification of the $s$--wave 
scattering length via Feshbach resonances has been demonstrated, been 
experimentally feasible the achievement of a quasi--ideal gas.

The main physical results of the paper are thus the following. 
We have been able to treat
analytically the quantum dynamics of an ideal gas of $N$ atom in the LD
limit. We have shown that the dynamics naturally splits into two parts:
a fast part, during which coherences between the degenerated states are preserved, and a slower part, during which thermal equilibrium is achieved.
Even though, the ideal case considered is not realistic
(at least without external modification of the scattering length), 
we think i) that it provides a lot of insight into more realistic situations; ii) it is, to our knowledge,
one of the extremely rare  examples of soluble quantum dynamical problems in the area of statistical physics. The method that we developed, and results can be carried over to more realistic situations concerning cooling of small atomic samples ($\simeq$20 atoms) in microtraps. Such situations are not far  from the reach of present experiments. The calculations for such a case should follow exactly the lines described in this paper, with the only difference that 
the parts of the Liouville equations describing the atom-atom interactions
that lift up the exact degeneracies should be included into the corrections
to the ${\cal L}^{(0)}$. In another words, they should be treated 
just like the corrections to the lowest order term in the LD expansion have been treated  in this paper. It is obvious, that as long as the splittings of the quasi-degenerated levels will remain small relative to $\hbar\omega$, such realistic system will exhibit basic effects presented in this paper, i.e.
step-wise dynamics on the two time scales.

The quantum dynamics of samples  of cold atoms exhibits, in our 
opinion, an enormous reachness of interesting physical and 
mathematical phenomena, such as multistable, exotic stationary 
states, multistage dynamics etc. The present paper is another 
example of this reachness. Further studies are, however, 
required to get more understanding of this new physics, 
including for instance  developement of other statistical 
physics tools \cite{kutner}(c), such as diffusion equations, 
hydrodynamic limits etc.

After this work was finished we have learned from T. Fischer, K. Vogel 
and W. Schleich
that similar algebra to the one used by us appears in the problem of 
the cooling of a sample of bosons with a simple particle reservoir 
\cite{Fischer}. We thank Yvan Castin, Jean Dalibard, Ralph Dum, 
T. Fischer, K. Vogel, and  P. 
Zoller for enlighting discussions. We acknowledge
the   support of the Deutsche Forschungsgemeinschaft (SFB
407), ESF PESC Proposal BEC2000+, and TMR ERBXTCT96-0002.

\begin{appendix}

\section{The structure of the Fock-Hilbert space}

The matrix elements of the operators $B$, and $C$ can be 
calculated using elegant algebraic methods. To this aim we 
first observe that the operators in question  fulfill the 
commutation relations
\begin{mathletters}
\begin{eqnarray}
\left[A,B\right]=& & 0, \label{pierh}\\
\left[A,C\right]=& & B/(2\sqrt{N}),\label{pierh1}\\
\left[A,B^{\dag}\right]=& & 2A^{\dag}/\sqrt{N} \label{drugh}, \\
\left[A,C^{\dag}\right]=& & \hat E/N+ 1/2 \label{drugh1}, \\
\left[B,B^{\dag}\right]=& &4\hat E/N+2,  \label{trzex}\\
\left[B,\hat E\right]=& &2B, \label{czwar}\\
\left[C,\hat E\right]=& &C, \label{czwar1}\\
\left[A,\hat E\right]=& &A, \label{czwar2}
\end{eqnarray}
\end{mathletters}
with $\hat E=\sum_{n=0}^{\infty} n a^{\dag}_na_n$ denoting the 
normalized energy operator.

It proves to be very useful to introduce the operator
\begin{equation}
D=B-A^2/\sqrt{N}.
\end{equation}
This operator fulfills 
\begin{mathletters}
\begin{eqnarray}
\left[D,A\right]=& & 0, \label{pierd}\\
\left[D,A^{\dag}\right]=& & 0\label{drughd}, \\
\left[D,D^{\dag}\right]=& & 4(\hat E- A^{\dag}A)/N+ 2(1-1/N)
 \label{drugh1d}, 
\end{eqnarray}
\end{mathletters}
Since the operators $A^{\dag}A$, and $D^{\dag}D$ commute, it is useful to
characterize the multiple vacua in terms of the eigenvalues of these
two hermitian operators.

Let $|0_{l0v}\rangle$ denote the states that fulfill
\begin{mathletters}
\begin{eqnarray}
A|0_{l0v}\rangle=0, \\
D|0_{l0v}\rangle=0.
\end{eqnarray}
\end{mathletters}
There are $m_N(l)$ such states, and the index $v$ enumerates them. 
Note that the states
\begin{equation}
|0_{lsv}\rangle= (D^{\dag})^s|0_{l-2s,0,v}\rangle/||(D^{\dag})^s|0_{l-2s,0,v}\rangle||,
\end{equation}
have energy $l$, are annihilated by $A$, and are eigenstates of $D^{\dag}D$  
with the  eigenvalue
\begin{equation}
\sum_{s'=0}^{s-1}\left[4(l-2s+2s')/N+2(1-1/N)\right].
\end{equation}
In the subsequent energy sectors we have thus the vacua: $|0_{001}\rangle$,
$|0_{211}\rangle$, $|0_{301}\rangle$, $|0_{421}\rangle$, $|0_{401}\rangle$,
$|0_{511}\rangle$, $|0_{501}\rangle$, $|0_{631}\rangle$, $|0_{611}\rangle$,
$|0_{601}\rangle$, $|0_{602}\rangle$, etc.

The Fock-Hilbert space is spanned by the vectors
\begin{equation}
|k_{lsv}\rangle=\frac{(A^{\dag})^k}{\sqrt{k!}}|0_{lsv}\rangle.
\end{equation}

\section{Construction of vacuum states}

We have seen in Appendix A that the vacuum states can be constructed by
 applying the operator $D^{\dag}$ consecutively to the states 
$|0_{l,0,v}\rangle$.
In this Appendix we present explicit construction of another family 
of the vacuum 
states that are annihilated by the operator $A$. In fact we 
consider a more general case with
\begin{equation}
  A=\sum_{n=0}A_n a_n^{\dag}a_{n+1}.
\label{defa1}
\end{equation}
Such defined operator reduces to the one given by Eq. (\ref{defa})
if we put $A_n=\sqrt{n+1}$. The vaccum states
fulfill
\begin{equation}
A|0_l\rangle= 0
\end{equation}

There is one obvious solution of the above equation which describes
the global ground state
\begin{equation}
|0_0\rangle = |N,0,0,\ldots\rangle.
\end{equation}

Apart from that, for $l=2,3,\ldots$ we define
\begin{eqnarray}
|0_l\rangle&&=\sum_{m=1}^{l-1} \alpha_m^l 
a^{\dag}_{l+1-m}|N-m,m-1,0, \ldots\rangle
\nonumber \\
&&+ \alpha_{l}^l |N-l,l,0, \ldots\rangle. 
\label{defva}
\end{eqnarray}
From the above definition it is clear, that different vacua are 
orthogonal. Applying $A$ to the above expression after 
elementary algebra we derive the recurrence formulas for the 
coefficients
\begin{equation}
\alpha^l_m=-\frac{A_0}{A_{l-m}}\sqrt{m}\sqrt{N-m}\alpha^l_{m+1}
\end{equation}
valid for $m=1,l-2$, and 
\begin{equation}
\alpha^l_{l-1}=-\frac{A_0}{A_{1}}\frac{\sqrt{l}\sqrt{N-l+1}}{\sqrt{l-1}}
\alpha^l_{l}.
\label{recutr2}
\end{equation}
From the above expression it is easy to construct explicitely 
the corresponding vacuum states. The value of $\alpha^l_l$ can 
be conveniently  chosen for normalisation of the states.

In the case considered in this paper ($A_n=\sqrt{n+1}$)
the first few normalized vacuum states are:
%\begin{mathletters}
\begin{eqnarray}
&&|0_2\rangle = \frac{1}{\sqrt{N}}\left(|N-2,2,0,\ldots\rangle\right.
\nonumber \\
&&-\left.\sqrt{N\!-\!1}|N-1,0,1,0,\ldots\rangle\right), 
\label{vac2}\\
&&|0_3\rangle = \frac{\sqrt{8}}{\sqrt{N^2\!+\!3N-2}}
\left(|N-3,3,0,\ldots\rangle\right.
\nonumber \\
&&-\left.\frac{\sqrt{3(N\!-\!2)}}{2}|N-2,1,1,0,\ldots\rangle\right. 
\nonumber \\
&&+\left.\frac{\sqrt{(N\!-\!1)(N\!-\!2)}}{2\sqrt{2}} 
|N-1,0,0,1,0,\ldots\rangle\right)
\label{vac3}\\
&&|0_4\rangle = \frac{3}{\sqrt{N^3\!-\!5N^2\!-\!3N\!+\!21}}
\left(|N-4,4,0,\ldots\rangle\right.
\nonumber \\
&&-\left.\frac{\sqrt{2(N\!-\!3)}}{3}|N-3,2,1,0,\ldots\rangle\right. 
\nonumber \\
&&+\left.\frac{2\sqrt{(N\!-\!2)(N\!-\!3)}}{3}|N-2,1,0,1,0,\ldots
\rangle\right.  
\nonumber \\
&&-\left.\frac{\sqrt{(N\!-\!1)(N\!-\!2)(N\!-\!3)}}{3}
|N-1,0,0,0,1,0,\ldots\rangle\right)
\label{vac4}
\end{eqnarray}
%\end{mathletters}
%
etc.

\section{Calculation of the matrix elements}

Let us first consider the operator B,
 and derive the explicit expressions for the martix elements
\begin{equation}
f_{lsv}=\langle 0_{l-2s-1v}|B|0_{lsv}\rangle=
\langle 0_{l-2s-1v}|D|0_{lsv}\rangle
\end{equation}
that enter Eq. (\ref{glowna}). Note that the coefficients $f_{lsv}$ 
are real, since the matrix elements of the operator $A$ in the 
non-interacting atom  basis are real (see Appendix A).  
Moreover, without any loss of generality $f_{lsv}$'s may be assumed 
to be non-negative. In the  following I will skip the index $v$ which
 is not affected by the dynamics.

Since we know that $B|0_{ls}\rangle\propto |0_{l-2s-1}\rangle$ for 
$l\ge 2s$,  we can write
\begin{equation}
B|0_{ls}\rangle =f_{ls}|0_{l-2s-1}\rangle.
\end{equation}
Similarly, using the commutation relation (\ref{drugh}) we can write
\begin{equation}
B^{\dag}|0_{l-2s-1v}\rangle ={f_{ls}}|0_{ls}\rangle + 
 \sqrt{2/N}|2_{l-2,s}\rangle,
\end{equation}
or
\begin{equation}
f_{ls}^2=\langle 0_{l-2,s-1}|BB^{\dag}|0_{l-2,s-1}\rangle -2/N.
\end{equation}

From the above expressions using the commutation relation (\ref{trzex})
we obtain
\begin{equation}
f_{l+2,+1}^2=f_{ls}^2 + 4l/N +2-2/N.
\end{equation}
The above recurrence can be easily solved yielding
\begin{eqnarray}
f_{ls}^2&&= (2-2/N)s +4((l-2s)s+s(s-1)) N .
\end{eqnarray}
since $f_{l-2s,0}=0$.

In general, we may write
\begin{equation}
B|k_{ls}\rangle = f_{ls}|k_{l-2,s-1}\rangle + \sqrt{k(k-1]/N}|(k-2)_{ls}
\rangle.
\label{bgene}
\end{equation}

 The above formulae provide a very efficient 
method of calculating all of the matrix 
elements of the operators $B$ and $B^{\dag}$.

It is a little more tedious to derive corresponding formulae 
for the operator $C$. From Eq. (\ref{pierh}) we obtain
\begin{equation}
AC|0_{ls}\rangle = \frac{1}{2\sqrt{N}}f_{ls}|0_{l-2,s-1}\rangle,
\end{equation}
so that
\begin{equation}
C|0_{ls}\rangle =  \frac{1}{2\sqrt{N}}f_{ls}|1_{l-2,s-1}\rangle 
+ \sum_{s'}^{E(l-1/2)}w_{ls'}|0_{l-1,s'}\rangle.
\end{equation}
The coefficients $w_{ls'}=\langle 0_{l-1,s-1}|C|0_{ls}\rangle$ may be also 
regarded to be real, and can be, for instance, determined 
directly from the definitions of the vacuum states. 

In general, we can write
\begin{eqnarray}
C|k_{ls}\rangle &=&  \frac{\sqrt{k+1}}{2\sqrt{N}}f_{ls}|k+1_{l-2,s-1}\rangle 
\nonumber \\
&+& \sum_{s'}^{E(l-1/2)}w_{ls'}|k_{l-1,s'}\rangle
+v_{ls}(k)|k-1_{ls}\rangle, \label{cgene}
\end{eqnarray}
with
\begin{equation}
v_{ls}(k+1)=\sqrt{\frac{k}{k+1}}v_{ls}(k) +\frac{1}{\sqrt{k+1}}
\left(\frac{k+l}{N}+\frac{1}{2}\right) ,
\end{equation}
and $v_{ls}(0)=0$. The above formulae allow for very efficient calculations 
of the matrix element of the operators $C$ and $C^{\dag}$, provided the states $|0_{m0v}\rangle$ are known.

The expression (\ref{cgene}) implies immediately that ${\cal 
L}_1^{(1)}$ does not contribute at all to the final equations 
(\ref{equar}) and (\ref{equar1}). Since the matrix $\rho$ is 
diagonal in the $k$ index, whereas the operators $A$ and 
$A^{\dag}$ change $k$ to $k-1$, and $k+1$ respectively, only 
those parts of the operators $C$ and $C^{\dag}$ that change 
$k\pm 1$ back to $k$ could contribute. It is evident from Eq. 
(\ref{cgene}), however, that these parts of $C$ and  $C^{\dag}$ 
do not change $lsv$. Therefore, their contributions to Eqs.  
(\ref{equar}) and (\ref{equar1}) vanish identically, due to the 
trace-like sums over $k$ appearing on the right hand side.

Similar considerations show that there is no mixing of the 
populations and the cumulative coherences in Eqs.  
(\ref{equar}) and (\ref{equar1}). Let us, for example, consider 
Eq.  (\ref{equar}) for the populations $n_{lsv}$. As in the 
previous case, the contributions of the parts of the operators 
$B$ and $B^{\dag}$ that do not change $lsv$ vanish identically, 
due to the trace-like sums over $k$ appearing on the right hand 
side. The parts of $B$ and $B^{\dag}$ that change $l$ by $\mp 
2$, and $s$ by $\mp 1$ do contribute, 
but they can only transform the parts of the 
density matrix proportional to $|k_{lsv}\rangle\langle k_{l's'v'}|$ 
into $\propto |k_{lsv}\rangle\langle k_{l's'v'}|$, or 
$\propto|k_{l\pm2,s\pm 1,v}\rangle\langle k_{l'\pm 2,s'\pm 1,v'}|$, i.e. they can 
only lead to couplings between the populations $n_{lsv}$ and 
$n_{l\pm 2,s\pm 1,v}$, or between the cumulative coherences $r_{lsv,l's'v'}$ 
and $r_{l\pm 2,s\pm 1,v, l'\pm 2, s'\pm 1,v'}$.

\section{Decay of coherences}

In this Appendix we keep a single index $w=(l,s,v)$, and denote
$w\pm 2=(l\pm 2,s\pm 1,v)$.
In order to prove that the cumulative coherences decay to zero 
on the slow time scale, we rewrite Eq. (\ref{glowna1}) in the 
form
\begin{equation}
\dot r_{ww'}=\dot r_{ww'}^{DB}+\dot r^{NEG}_{ww'},
\label{kinto}
\end{equation}
where the first term
\begin{eqnarray}
&&\dot r^{DB}_{ww'} = 2\Gamma^{(1)}_{\downarrow}\left[f_{w+2}f_{w'+2} 
r_{w+2w'+2}
- f_wf_{w'}r_{ww'}\right] \nonumber \\
&&+ 2\Gamma^{(1)}_{\uparrow}\left[f_wf_{w'} r_{w-2w'-2}
- f_{w+2}f_{w'+2}r_{ww'}\right],\label{glownaDB}
\end{eqnarray}
corresponds to a set of kinetic equations with (positive) rates 
that fulfill detailed balance conditions. The matrix that 
enters the right hand side and generates the evolution is 
therefore evidently non-positively defined, and has exactly one 
eigenvector with zero eigenvalue, corresponding to   the 
stationary solution of the Boltzmann-Gibbs form.

The second term in Eq. (\ref{kinto})
\begin{eqnarray}
&&\dot r^{NEG}_{ww'} = -2\Gamma^{(1)}_{\downarrow}
\left[  \frac{1}{2}\left(f_w^2 +f_{w'}^2 -f_{w}f_{w'}
\right)r_{ww'}\right] \nonumber \\
&&-2\Gamma^{(1)}_{\uparrow}\left[ \frac{1}{2}
\left(f_{w+2}^2 +f_{w'+2}-f_{w+2}f_{w'+2}
\right)r_{ww'}\right],\label{glownaNEG}
\end{eqnarray}
corresponds to a set of simple decay equations, with the rates 
which are evidently positive, since $ \frac{1}{2}(f_w^2 
+f_{w'}^2)-f_{w}f_{w'}$ is strictly greater than zero for $w\ne 
w'$. 

The full dynamics of the cumulative coherences is thus 
generated by the sum of the two martices, one of which is 
non-positively defined, and the other being strictly negatively 
defined. The sum itself must therefore   be negatively defined, 
{\it ergo} it generates the decay to zero.

\end{appendix}

% -------------------------------------------------------------

\end{multicols}


\begin{references}
\bibitem{fotka} For instance, the statistics of photons emitted 
in the laser cooling process might be directly obtained from 
the statistics of the jumps between the states of the system, 
see \cite{CLZ}.
\bibitem{garzol}  It is also worth stressing there are 
attempts to describe the standard
evaporative cooling process in terms of a master equation, i.e. dividing 
a system of the same atoms into a proper subsystem and  a "bath". This is done by projecting the density matrix onto the "relevant" atomic degrees of freedom
using a coarse-graining procedure \cite{QKpapers}. 
Although in principle  our phenomenological
"bath" description can be applied in such a case, 
we shall not discuss it in this paper. Evaporative 
cooling involves typically several thousands of atoms, 
when atom-atom interaction contribute significantly, if not
dominantly to the dynamics of the system.
\bibitem{noteii} 
Rigorously, for $|\alpha|$ sufficiently large, the condition (\ref{conld}) 
cannot be satisfied. Note, however, that this corresponds to 
transitions {\it from}, or {\it to} very excited levels. 
The transitions {\it from} the very excited levels can be excluded from 
the begining, if the atomic sample is already at some low 
temperature, since those levels will  not be occupied. 
The transitions {\it to} excited levels, on the other hand,  are also not possible, if 
the bath is at sufficiently low temperature, since the function 
$n(k)$ provides an effective cut--off of the corresponding rates 
[see Eq. (\ref{crates2})].
\end{references}
\end{document}